\documentclass[onecolumn,nofootinbib]{revtex4}
\usepackage{graphicx}
\usepackage{latexsym}

\begin{document}

\title{Dark Energy Constraints after Planck}

\author{Jun-Qing Xia${}^{a}$}
\author{Hong Li${}^{a,b}$}
\author{Xinmin Zhang${}^{c}$}

\affiliation{${}^a$Key Laboratory of Particle Astrophysics, Institute of High Energy Physics, Chinese Academy of Science, P. O. Box 918-3, Beijing 100049, P. R. China}
\affiliation{${}^b$National Astronomical Observatories, Chinese Academy of Sciences, Beijing 100012, P. R. China}
\affiliation{${}^c$Theoretical Physics Division, Institute of High Energy Physics, Chinese Academy of Science, P. O. Box 918-4, Beijing 100049, P. R. China}


\begin{abstract}

The Planck collaboration has recently published maps of the Cosmic Microwave Background radiation with the highest precision. In the standard flat $\Lambda$CDM framework, Planck data show that the Hubble constant $H_0$ is in tension with that measured by the several direct probes on $H_0$. In this paper, we perform a global analysis from the current observational data in the general dark energy models and find that resolving this tension requires the dark energy model with its equation of state (EoS) $w\neq-1$. Firstly, assuming the $w$ to be a constant, the Planck data favor $w < -1$ at about $2\,\sigma$ confidence level when combining with the supernovae ``SNLS'' compilation. And consequently the value derived on $H_0$, $H_0=71.3\pm2.0$ ${\rm km\,s^{-1}\,Mpc^{-1}}$ ($68\%$ C.L.), is consistent with that from direct $H_0$ probes. We then investigate the dark energy model with a time-evolving $w$, and obtain the $68\%$ C.L. constraints $w_0=-0.81\pm0.19$ and $w_a=-1.9\pm1.1$ from the Planck data and the ``SNLS'' compilation. Current data still slightly favor the Quintom dark energy scenario with EoS across the cosmological constant boundary $w\equiv-1$.

\end{abstract}


\maketitle

\section{Introduction}\label{int}

With the accumulation of observational data from cosmic microwave background measurements (CMB), large scale structure surveys (LSS), and type Ia supernovae observations (SNIa) and the improvements of the data quality, the cosmological observations play a crucial role in our understanding of the universe and also in constraining the cosmological parameters. Recently, the Planck collaboration has released the first cosmological papers providing the highest resolution, full sky, maps of the CMB temperature anisotropies \cite{planck_map}. The corresponding numerical analysis of cosmological parameters indicates the concordance cosmology in which our Universe is flat and currently dominated by dark matter and dark energy \cite{planck_fit}.

Due to the improved precision, this new Planck data has constrained several cosmological parameters at few percent level. It is crucial to compare these constraints from different combinations of cosmological datasets and check whether they are in agreement with each other. In the standard flat $\Lambda$CDM framework, the constraint on the Hubble constant $H_0$ is significantly improved by the new Planck data, namely $H_0=67.4\pm1.4$ ${\rm km\,s^{-1}\,Mpc^{-1}}$ at $68\%$ confidence level \cite{planck_fit}. However, this result is obviously in tension with that measured by various lower-redshift methods, such as the direct $H_0$ probe from the Hubble Space Telescope (HST) observations of Cepheid variables with $H_0=73.8\pm2.4$ ${\rm km\,s^{-1}\,Mpc^{-1}}$ \cite{hst_riess} or $H_0=74.3\pm1.5{\rm (stat.)}\pm2.1{\rm (sys.)}$ ${\rm km\,s^{-1}\,Mpc^{-1}}$ \cite{hst_freedman}. If this difference is not induced by the unknown systematics in measurements, the tension between measurements of $H_0$ could imply that the concordance cosmological model (the flat $\Lambda$CDM model) is in fact incomplete. Therefore, it is very important to perform the global analysis again and extend the simple assumption to some more general cosmological models to see whether this tension could be resolved \cite{verde,waynehu,huang}.

In this paper, we extend the $\Lambda$CDM to the dynamical dark energy models and explore the cosmological constraints on the equation of state of dark energy and the Hubble constant from the latest cosmological data sets, including the Planck and also the WMAP9 data for comparison, the BAO measurements from several LSS surveys, the ``SNLS'' compilation which includes 473 supernovae reprocessed by Ref.\cite{snls} and the HST gaussian prior on the Hubble constant $H_0$ given by Ref.\cite{hst_riess}. Firstly, we consider the dark energy model with a constant equation of state (EoS) $w$ and find that the current data favor the model with $w<-1$ at about $2\,\sigma$ confidence level. Due to the strong anti-correlation between $w$ and $H_0$, the constraint on $H_0$ is now consistent with the HST direct probe and the $H_0$ tension disappears. Then we focus on the time-evolving EoS ($w(z)$) dark energy model. We compare our numerical method provided in Refs.\cite{pert_zhao,pert_xia} to treat the dark energy perturbations consistently in the whole parameter space with the method given by Ref.\cite{pert_fang} and find that the constraints on dark energy parameters with these two methods are almost identical. Finally, based on our method we constrain the time-evolving EoS dark energy model and find that current data still slightly favor the Quintom dark energy scenario whose EoS can cross the cosmological constant boundary $w\equiv-1$.

Our paper is organized as follows: In Section \ref{data} we describe the method and the latest observational data sets used in the numerical analyses; Section \ref{tension} and section \ref{quintom} contains our main global constraints on the EoS of dark energy models and the Hubble constant from the current observations. The last Section \ref{sum} is the conclusions.

\section{Method and Data}\label{data}

\subsection{Numerical Method}

We perform a global fitting of cosmological parameters using the {\tt CosmoMC} package \cite{cosmomc}, a Markov Chain Monte Carlo (MCMC) code. We assume purely adiabatic initial conditions and neglect the primordial tensor fluctuations. The basic six cosmological parameters are allowed to vary with top-hat priors: the cold dark matter energy density parameter $\Omega_c h^2 \in [0.01, 0.99]$, the baryon energy density parameter $\Omega_b h^2 \in [0.005, 0.1]$, the scalar spectral index $n_s \in [0.5, 1.5]$, the primordial amplitude $\ln[10^{10}A_s] \in [2.7, 4.0]$, the ratio (multiplied by 100) of the sound horizon at decoupling to the angular diameter distance to the last scattering surface $100\Theta_s \in [0.5, 10]$, and the optical depth to reionization $\tau \in [0.01, 0.8]$. The pivot scale is set at $k_{s0} = 0.05$ ${\rm Mpc}^{-1}$.

For dark energy models with a general w(z), we choose the popular parametrization of EoS given by Refs.\cite{cpl}:
\begin{equation}\label{eq_cpl}
w_{\rm de}(a) = w_0 + w_a (1-a)~,
\end{equation}
where $a\equiv 1/(1+z)$ is the scale factor and $w_a = - dw/da$ characterizes the ``running'' of EoS, and set the top-hat priors $w_0 \in [-2,0]$ and $w_a \in [-10,2]$. For the standard $\Lambda$CDM model it corresponds to $w_0\equiv-1$ and $w_a \equiv 0$. And for the model with a constant EoS ($w$CDM), it equals to taking $w_a \equiv 0$. When using the global fitting strategy to constrain the cosmological parameters, it is crucial to include dark energy perturbations \cite{pert_zhao}. In section \ref{perturbation} we will discuss the method provided in Refs.\cite{pert_zhao,pert_xia} in detail. Therefore, the most general parameter space in our analyses is:
\begin{equation}
\left\{ \Omega_bh^2, \Omega_ch^2, \Theta_s, \tau, n_s, A_s, w_0, w_a \right\}~.
\end{equation}

\subsection{Current Observational Data}

In our analysis, we consider the following cosmological probes: i) power spectra of CMB temperature and polarization anisotropies; ii) the baryon acoustic oscillation in the galaxy power spectra; iii) measurement of the current Hubble constant; iv) luminosity distances of type Ia supernovae.

For the Planck data from the 1-year data release \cite{planck_fit}, we use the low-$\ell$ and high-$\ell$ CMB temperature power spectrum data from Planck with the low-$\ell$ WMAP9 polarization data (Planck+WP). We marginalize over the nuisance parameters that model the unresolved foregrounds with wide priors \cite{planck_likelihood}, and do not include the CMB lensing data from Planck \cite{planck_lens}. For comparison, we also use the WMAP9 CMB temperature and polarization power spectra \cite{wmap9} in our calculations.

Baryon Acoustic Oscillations provides an efficient method for measuring the expansion history by using features in the clustering of galaxies within large scale surveys as a ruler with which to measure the distance-redshift relation. It provides a particularly robust quantity to measure \cite{bao}. It measures not only the angular diameter distance, $D_A(z)$, but also the expansion rate of the universe, $H(z)$, which is powerful for studying dark energy \cite{task}. Since the current BAO data are not accurate enough for extracting the information of $D_A(z)$ and $H(z)$ separately \cite{okumura}, one can only determine an effective distance \cite{baosdss}:
\begin{equation}
D_V(z)=[(1+z)^2D_A^2(z)cz/H(z)]^{1/3}~.
\end{equation}
Following the Planck analysis \cite{planck_fit}, in this paper we use  the BAO measurement from the 6dF Galaxy Redshift Survey (6dFGRS) at a low redshift ($r_s/D_V (z = 0.106) = 0.336\pm0.015$) \cite{6dfgrs}, and the measurement of the BAO scale based on a re-analysis of the Luminous Red Galaxies (LRG) sample from Sloan Digital Sky Survey (SDSS) Data Release 7 at the median redshift ($r_s/D_V (z = 0.35) = 0.1126\pm0.0022$) \cite{sdssdr7}, and the BAO signal from BOSS CMASS DR9 data at ($r_s/D_V (z = 0.57) = 0.0732\pm0.0012$) \cite{sdssdr9}.

In our analysis, we add a Gaussian prior on the current Hubble constant given by Ref.\cite{hst_riess}; $H_0 = 73.8 \pm 2.4$ ${\rm km\,s^{-1}\,Mpc^{-1}}$ (68\% C.L.). The quoted error includes both statistical and systematic errors. This measurement of $H_0$ is obtained from the magnitude-redshift relation of 240 low-z Type Ia supernovae at $z < 0.1$ by the Near Infrared Camera and Multi-Object Spectrometer (NICMOS) Camera 2 of the Hubble Space Telescope (HST). This is a significant improvement over the previous prior, $H_0 = 72 \pm 8$ ${\rm km\,s^{-1}\,Mpc^{-1}}$, which is from the Hubble Key project final result. In addition, we impose a weak top-hat prior on the Hubble parameter: $H_0 \in [40, 100]$ ${\rm km\,s^{-1}\,Mpc^{-1}}$.

Finally, we include data from Type Ia supernovae, which consists of luminosity distance measurements as a function of redshift, $D_L(z)$. In this paper we use the supernovae data set, ``SNLS'' compilation, which includes 473 supernovae reprocessed by Ref.\cite{snls}. When calculating the likelihood, we marginalize the nuisance parameters, like the absolute magnitude $M$ and the parameters $\alpha$ and $\beta$, as explained by Ref. \cite{planck_fit}.

\section{Dark Energy with Constant $w$}\label{tension}

In this section we re-examine the possible tension between constraints on $H_0$ from Planck and the local direct $H_0$ probes in the $\Lambda$CDM and $w$CDM models and list the constraints on some related cosmological parameters from different data combinations in Table \ref{wcdmtable}.

Firstly, we consider the flat $\Lambda$CDM model. When using the CMB data alone, we obtain $68\%$ C.L. constraints on the Hubble constant: $H_0=69.2\pm2.2$ ${\rm km\,s^{-1}\,Mpc^{-1}}$ and $H_0=67.0\pm1.2$ ${\rm km\,s^{-1}\,Mpc^{-1}}$ from WMAP9 and Planck+WP, respectively. In order to quantify the difference of constraints on parameter $X$ obtained from two probes $A$ and $B$, here we simply define a variable $T(X)$:
\begin{equation}
T(X) \equiv \frac{|P_{X,A} - P_{X,B}|}{\sigma_{X,A}}~,
\end{equation}
where $P$ and $\sigma$ are the median value and the standard error bar of parameter $X$, respectively. In this case, the difference variable $T(H_0)=1.8$, which means the median value of $H_0$ from Planck+WP departs from that obtained from WMAP9 at about $1.8\,\sigma$ confidence level. This $2.2$ ${\rm km\,s^{-1}\,Mpc^{-1}}$ shift in $H_0$ between Planck+WP and WMAP9 is mainly due to the slightly higher matter density determined by Planck+WP ($\Omega_mh^2=0.143\pm0.003$ ($1\,\sigma$)) compared to WMAP9 ($\Omega_mh^2=0.136\pm0.004$ ($1\,\sigma$)). Since the Hubble constant is strongly anti-correlated with the current matter density fraction $\Omega_m$, we obtain the $68\%$ C.L. limits on the current matter density fraction of $\Omega_m=0.289\pm0.027$ and $\Omega_m=0.319\pm0.017$ from WMAP9 and Planck+WP data, separately, namely the difference $T(\Omega_m)=1.7$ which is similar with that of the Hubble constant. Besides the WMAP9 data, we also compare the constraints of $H_0$ from Planck+WP and the HST prior. Due to small error bars and the large discrepancy between their median values, the tension becomes larger, $T(H_0)=5.7$, which is found by the Planck data \cite{planck_fit}. This huge discrepancy cannot be easily resolved by varying the parameters of the standard $\Lambda$CDM model.

Then, we move to the $w$CDM framework to see whether the tension between constraints on $H_0$ would be relaxed.  The CMB anisotropies mainly contain the information about the high-redshift universe, but it is not directly sensitive to phenomena which affect the lower redshift Universe, such as the nature of dark energy. There are very strong degeneracies among $w$, $\Omega_m$ and $H_0$ \cite{Li:2012ug}. Therefore, the CMB data alone now can not able to constrain them very well, namely their $95\%$ limits of $H_0$ and $w$ are $H_0=71\pm15$ ${\rm km\,s^{-1}\,Mpc^{-1}}$, $w=-1.03\pm0.46$ and $H_0=83\pm11$, $w=-1.50\pm0.31$ from WMAP9 and Planck+WP data alone, respectively. These constraints are also consistent with the HST prior, due to the large error bars. In order to break these degeneracies, inferring $w$ or $H_0$ from CMB data requires the combination with lower redshift probes.

\begin{table}
\caption{The median values and $1\,\sigma$ error bars on some cosmological parameters obtained from different data combinations in different dark energy models. In each data combination, we also list the difference between the minimal $\chi^2$ obtained in the $w$CDM and $\Lambda$CDM models, $\Delta\chi^2_{\rm min}\equiv\chi^2_{\rm min}(w{\rm CDM})-\chi^2_{\rm min}(\Lambda{\rm CDM})$.}\label{wcdmtable}
\begin{center}

\begin{tabular}{c|c|c|c|c|c|c}

\hline\hline
Dark Energy Models&\multicolumn{2}{|c|}{$\Lambda$CDM}&\multicolumn{4}{|c}{$w$CDM}\\
\hline
Parameters&$H_0$&$\Omega_m$&$H_0$&$\Omega_m$&$w$&$\Delta\chi_{\rm min}^2$\\
\hline
WMAP9 alone&$69.2\pm2.2$&$0.289\pm0.027$&$71\pm15$&$0.312\pm0.120$&$-1.031\pm0.460$&$-0.13$\\
WMAP9+BAO&$68.3\pm1.0$&$0.298\pm0.012$&$68.5\pm3.1$&$0.297\pm0.021$&$-1.012\pm0.150$&$-0.02$\\
WMAP9+SNLS&$70.6\pm1.9$&$0.270\pm0.021$&$71.8\pm2.4$&$0.266\pm0.022$&$-1.072\pm0.078$&$-0.87$\\
WMAP9+HST&$71.3\pm1.6$&$0.264\pm0.018$&$74.0\pm2.5$&$0.252\pm0.019$&$-1.153\pm0.120$&$-1.87$\\
WMAP9+SNLS+HST&$71.8\pm1.6$&$0.258\pm0.017$&$72.9\pm1.8$&$0.257\pm0.016$&$-1.092\pm0.071$&$-1.96$\\
\hline
Planck+WP&$67.0\pm1.2$&$0.319\pm0.017$&$83\pm11$&$0.216\pm0.061$&$-1.501\pm0.310$&$-2.96$\\
Planck+WP+BAO&$67.7\pm0.8$&$0.309\pm0.011$&$70.8\pm2.9$&$0.287\pm0.021$&$-1.142\pm0.120$&$-0.87$\\
Planck+WP+SNLS&$67.9\pm1.1$&$0.307\pm0.015$&$71.3\pm2.0$&$0.281\pm0.018$&$-1.137\pm0.067$&$-4.48$\\
Planck+WP+HST&$68.4\pm1.1$&$0.300\pm0.014$&$74.6\pm2.6$&$0.258\pm0.018$&$-1.250\pm0.096$&$-6.75$\\
Planck+WP+SNLS+HST&$68.9\pm1.0$&$0.293\pm0.013$&$72.5\pm1.7$&$0.270\pm0.014$&$-1.167\pm0.061$&$-8.82$\\

\hline\hline
\end{tabular}
\end{center}
\end{table}

\begin{figure}[t]
\begin{center}
\includegraphics[scale=0.44]{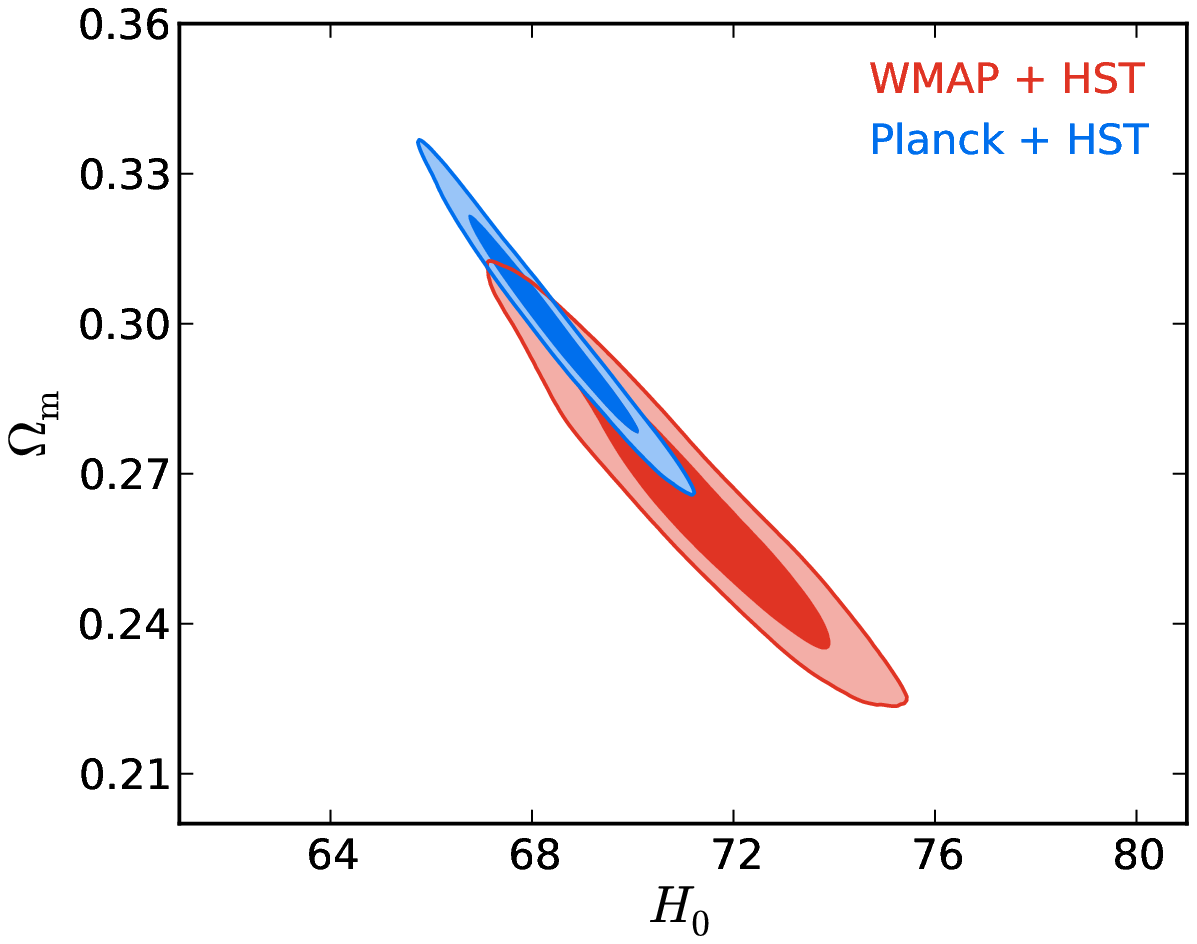}
\includegraphics[scale=0.44]{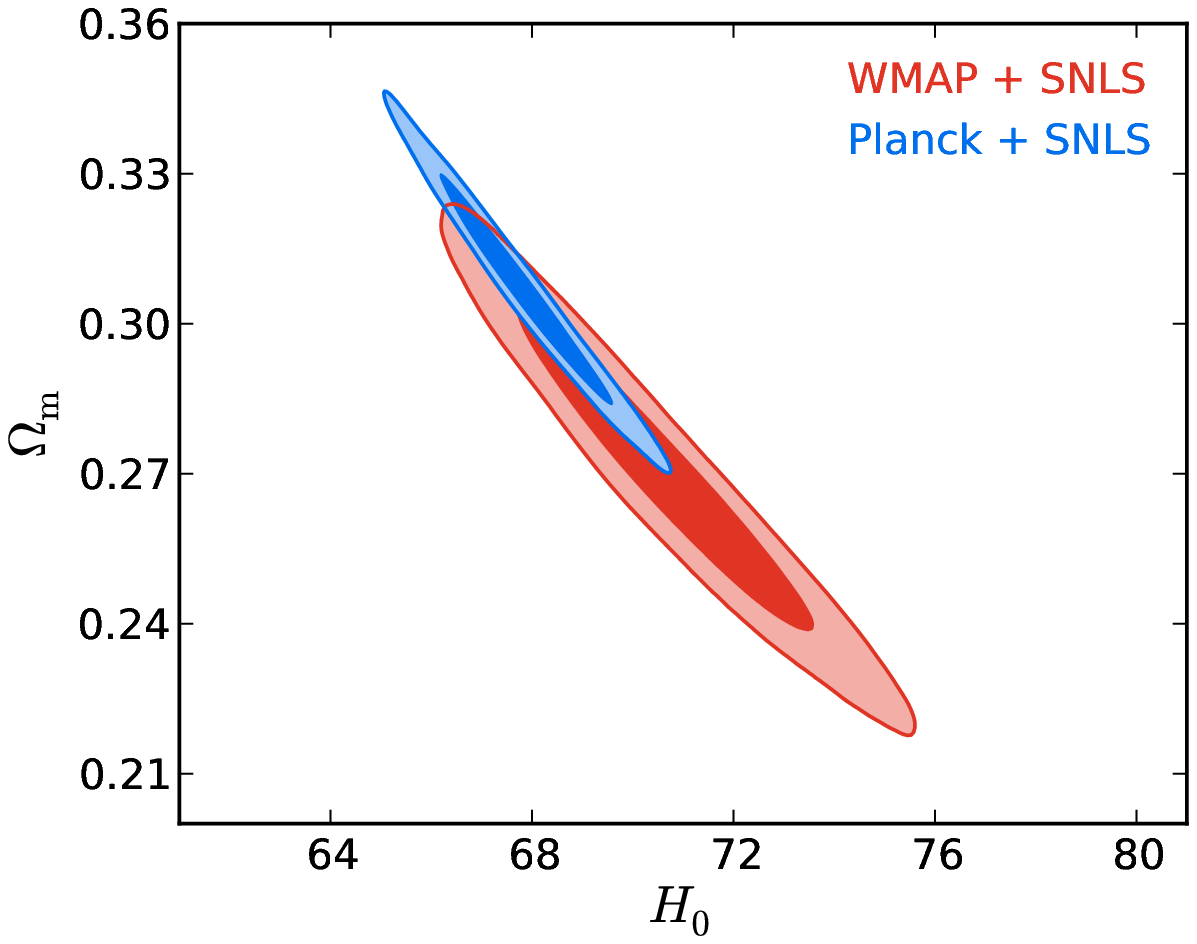}
\includegraphics[scale=0.44]{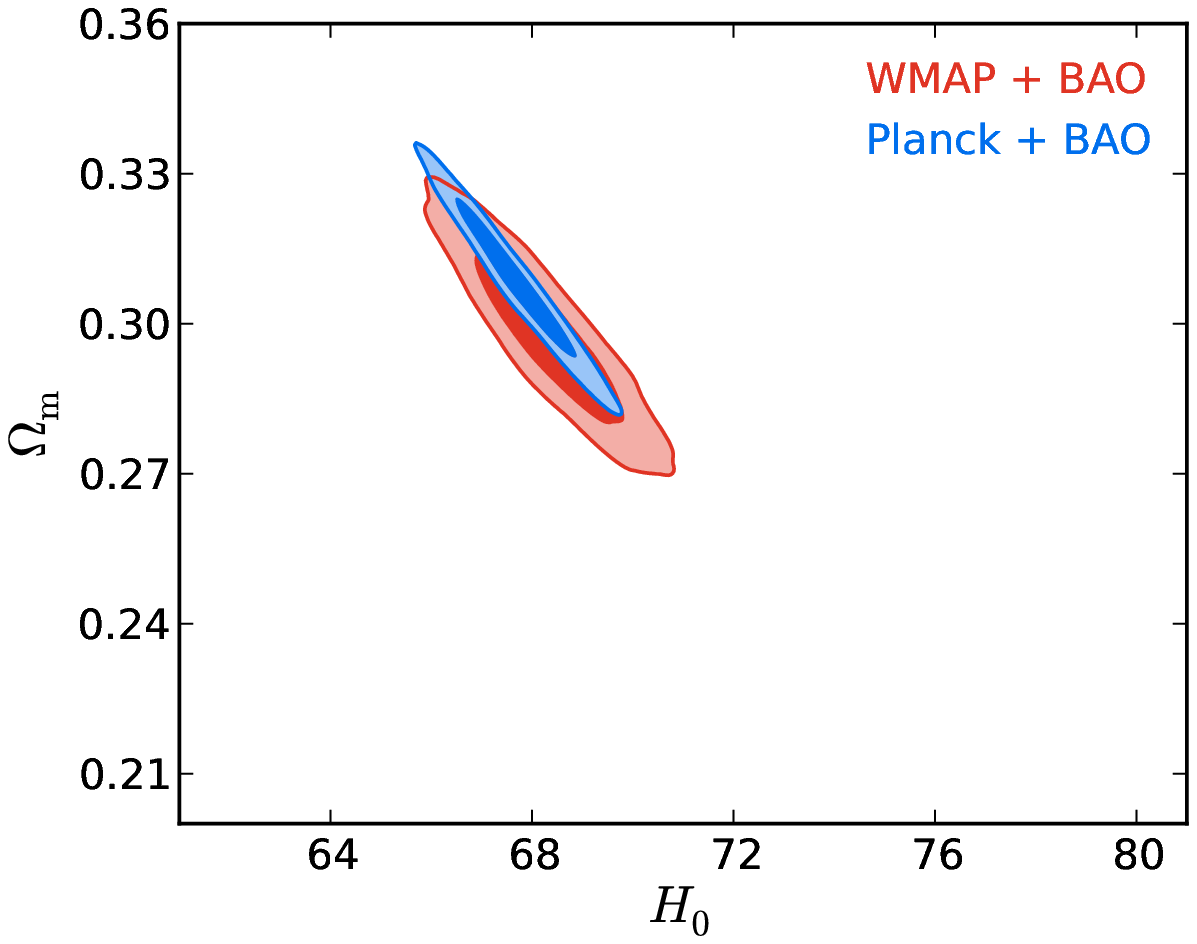}
\includegraphics[scale=0.44]{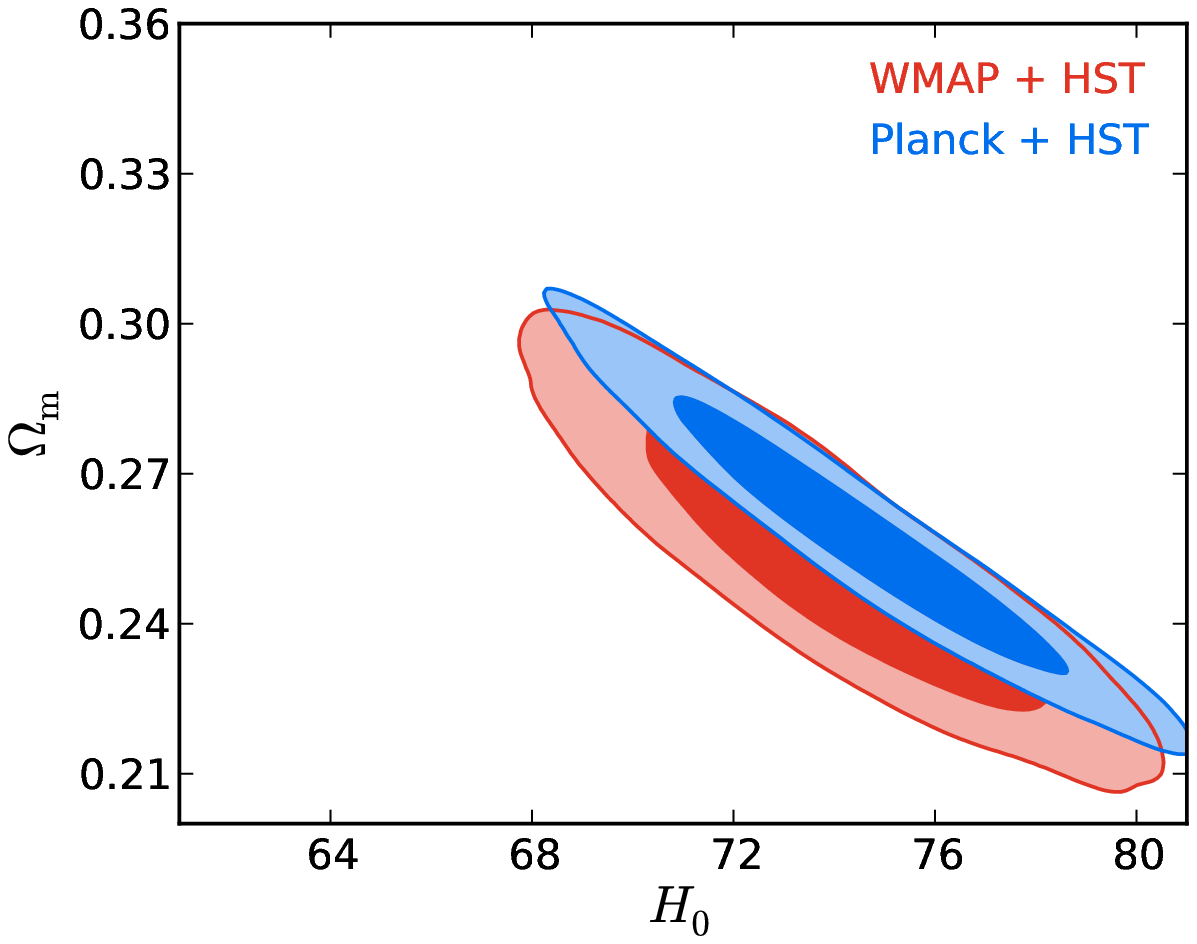}
\includegraphics[scale=0.44]{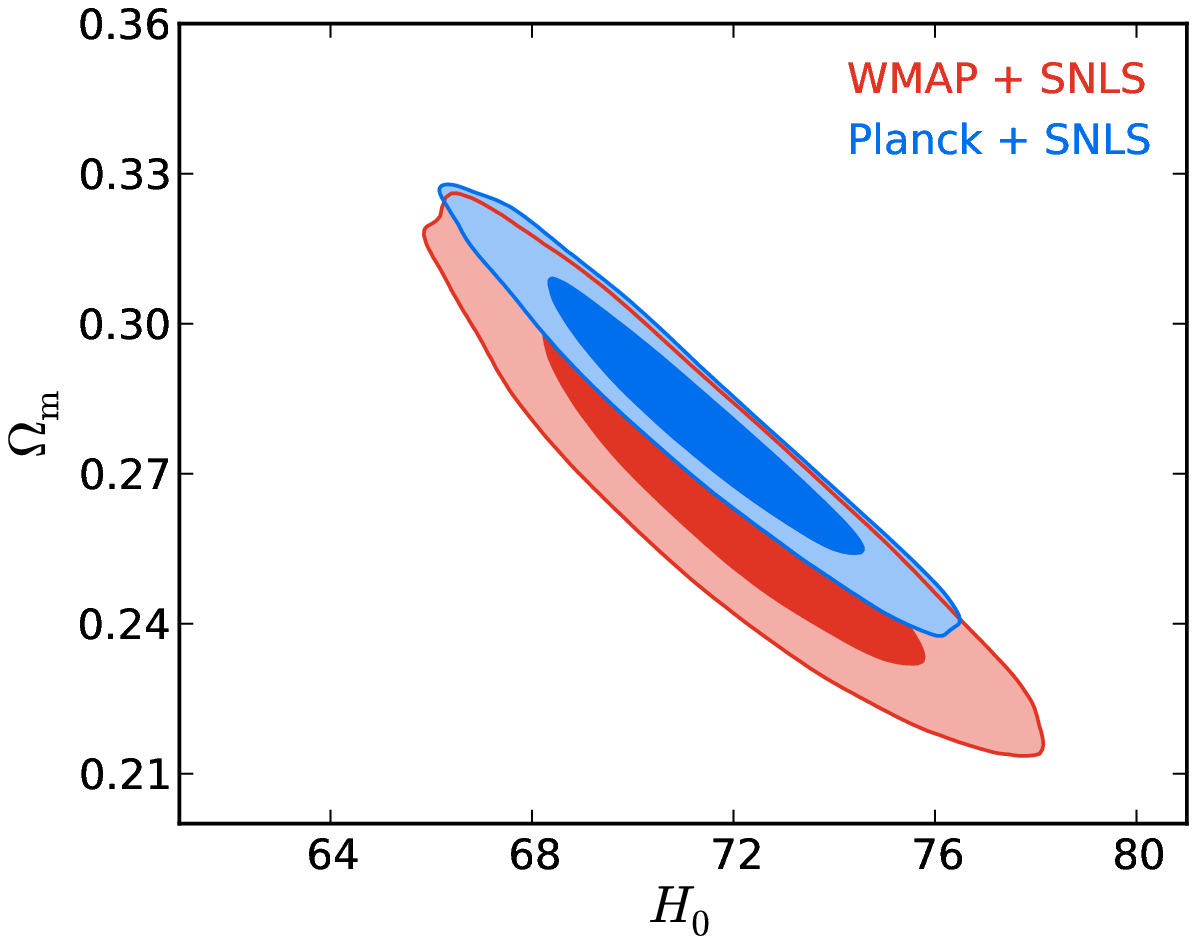}
\includegraphics[scale=0.44]{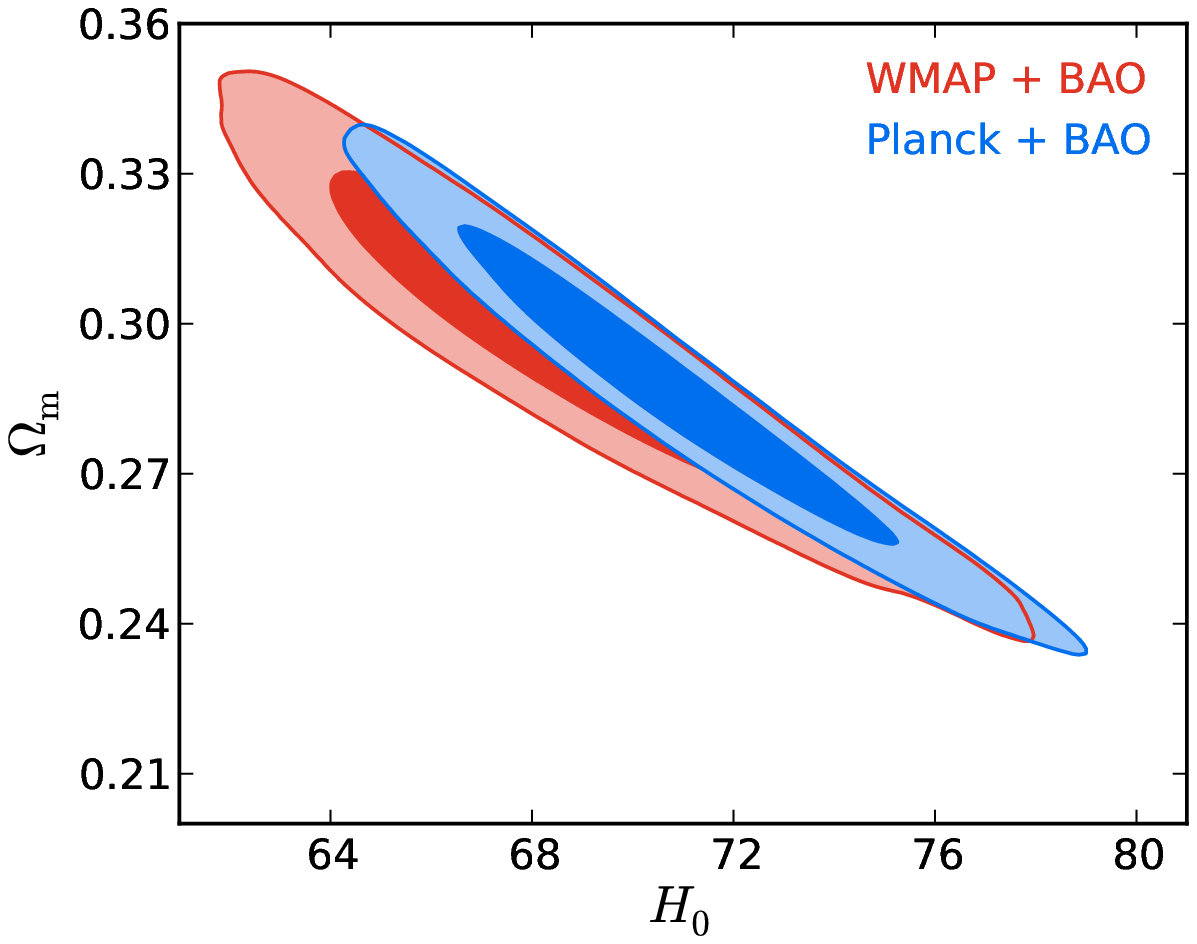}
\caption{Marginalized two-dimensional likelihood (1, $2\,\sigma$ contours) constraints on the parameters $H_0$ and $\Omega_m$ in the $\Lambda$CDM model (upper panels) and the $w$CDM model (lower panels) from different data combinations. Red and blue contours denote the data combinations with WMAP9 and with Planck+WP, respectively. \label{wcdm}}
\end{center}
\end{figure}

We first add the HST gaussian prior on $H_0$. In the $\Lambda$CDM model, this HST prior does not affect the constraint from WMAP9 too much: $H_0=71.3\pm1.6$ ${\rm km\,s^{-1}\,Mpc^{-1}}$ ($68\%$ C.L.), since the constraint on $H_0$ from WMAP9 alone is consistent with the HST prior. In contrast, due to the large discrepancy on $H_0$ between Planck+WP and HST prior, adding the HST prior to the Planck+WP data forces the obtained median value of $H_0$ towards to the higher one, namely $H_0=68.4\pm1.1$ ${\rm km\,s^{-1}\,Mpc^{-1}}$ ($68\%$ C.L.). The difference between constraints on $H_0$ from Planck+WP+HST and WMAP9+HST or the HST prior are still very large, $T(H_0)=2.6$ and $T(H_0)=4.9$, respectively. Consequently, the constraints on $\Omega_m$ from WMAP9+HST and Planck+WP+HST are also quite different, which is clearly shown in the first upper panel of Figure \ref{wcdm}. When including the constant dark energy EoS, the constraints on $H_0$ and $\Omega_m$ from Planck+WP+HST and WMAP9+HST, shown in the first lower panel of Figure \ref{wcdm}, are quite similar. Planck+WP+HST data combination gives $H_0=74.6\pm2.6$ ${\rm km\,s^{-1}\,Mpc^{-1}}$ and $\Omega_m=0.258\pm0.018$ in the $w$CDM framework, which is in good agreement with those from WMAP9 and HST prior. In this case, the $68\%$ C.L. constraint on $w$ from Planck+WP+HST is $w=-1.250\pm0.096$, which deviates from $w\equiv-1$ at more than $2\,\sigma$ confidence level.

In this case, the minimal $\chi^2$ from Planck+WP+HST becomes smaller in the $w$CDM model than that obtained in the standard $\Lambda$CDM model, namely $\Delta\chi_{\rm min}^2=-6.75$, which is shown in Table \ref{wcdmtable}. Based on the akaike information criterion (AIC):
\begin{equation}
{\rm AIC}\equiv-2\ln{\mathcal{L}_{\rm max}}+2k~,
\end{equation}
where $\mathcal{L}_{\rm max}$ is the maximum likelihood achievable by the model and $k$ the number of parameters of the model \cite{aic}, we obtain the difference on the AIC between the $\Lambda$CDM and $w$CDM, $\Delta{\rm AIC}\equiv{\rm AIC}(w{\rm CDM})-{\rm AIC}(\Lambda{\rm CDM})=-4.75$. The $w$CDM model with one more free parameter $w$ is strongly favored by the Planck+WP+HST data combination.

Then, we include the SNLS supernovae sample into the calculations and find the similar conclusion with that obtained by adding the HST prior. In the $\Lambda$CDM model, the constraint on $H_0$ from Planck+WP+SNLS is in tension with that from WMAP9+SNLS or the HST prior, and the discrepancies are about $2.5\,\sigma$ and $5.4\,\sigma$, respectively. However, the $w$CDM dark energy model resolves this tension and the constraint on $w$ is $w=-1.137\pm0.067$ at $68\%$ confidence level. The differences on the minimal $\chi^2$ and the AIC from the $w$CDM and $\Lambda$CDM models are $\Delta\chi^2_{\rm min}=-4.48$ and $\Delta{\rm AIC}=-2.48$. The Planck+WP+SNLS data also favor the $w$CDM model obviously.

\begin{figure}[t]
\begin{center}
\includegraphics[scale=0.6]{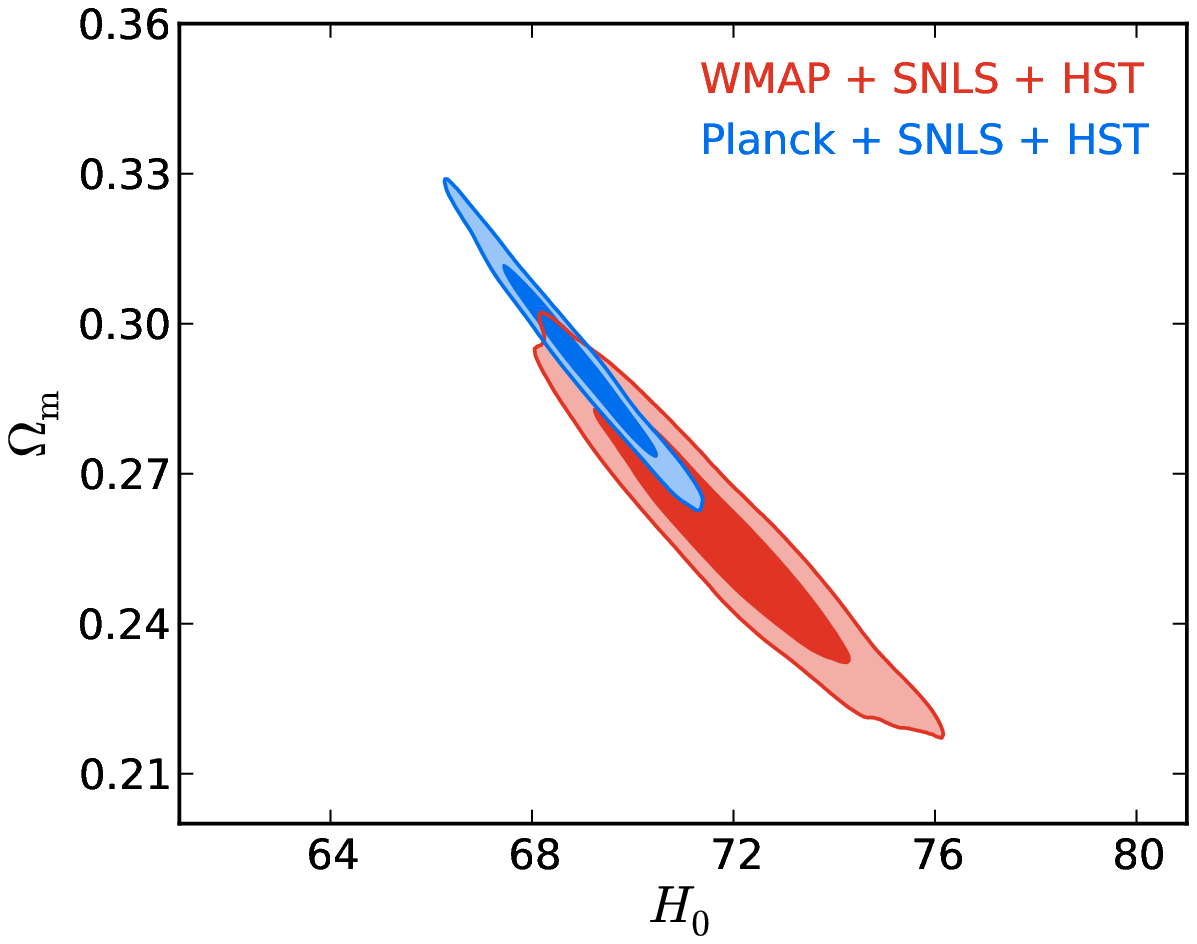}
\includegraphics[scale=0.6]{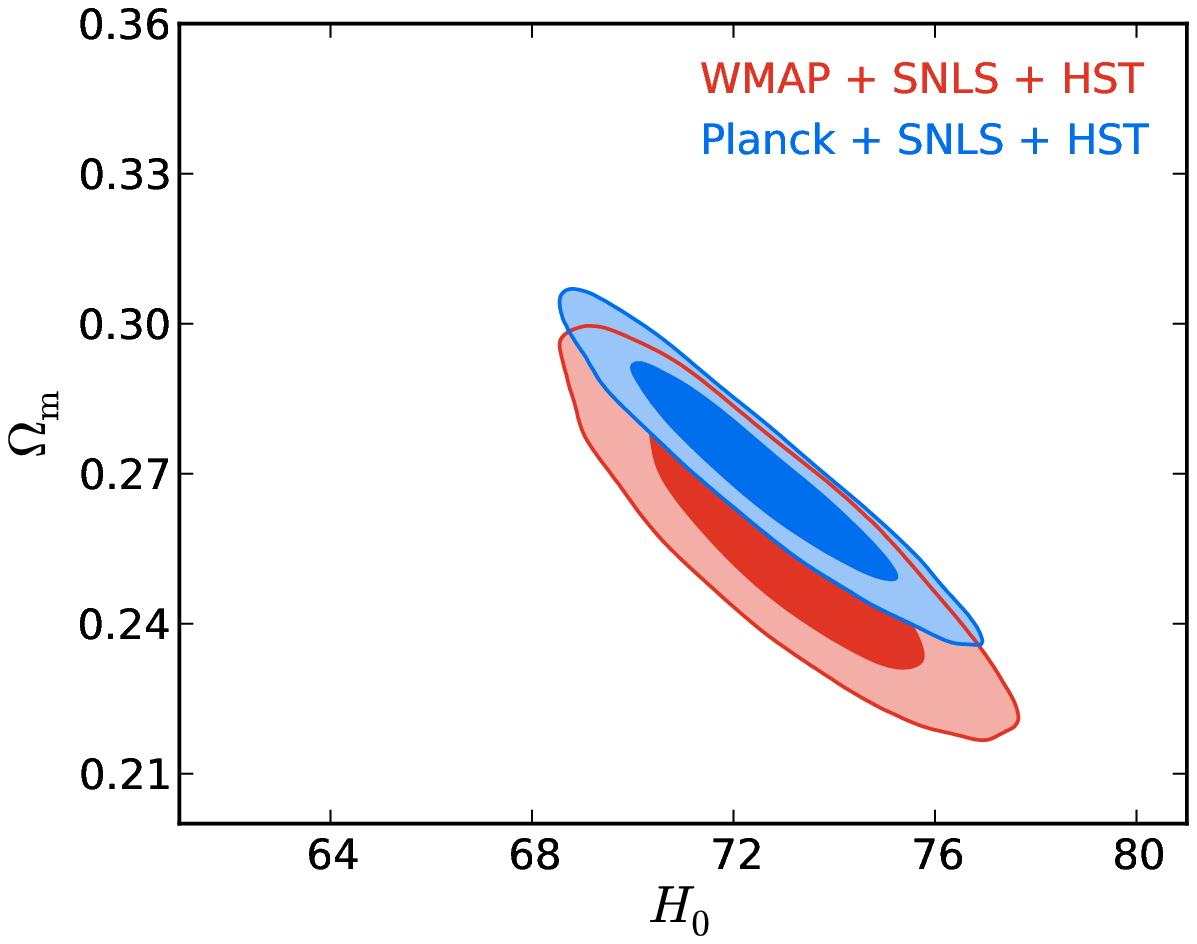}
\caption{Marginalized two-dimensional likelihood (1, $2\,\sigma$ contours) constraints on the parameters $H_0$ and $\Omega_m$ in the standard $\Lambda$CDM model (left panel) and the $w$CDM model (right panel) from different data combinations. Red and blue contours denote the data combinations with WMAP9 and with Planck+WP, respectively. \label{wcdm_all}}
\end{center}
\end{figure}

We also combine the SNLS and the HST prior together, the situation becomes worse in the $\Lambda$CDM model, shown in the left panel of Figure \ref{wcdm_all}. The Planck data combination yields the 68\% C.L. limit on the Hubble constant of $H_0=68.9\pm1.0$ ${\rm km\,s^{-1}\,Mpc^{-1}}$, which departs from the values from WMAP9 data combination or the HST prior at more than $2.9\,\sigma$ and $4.4\,\sigma$ confidence level. When varying the dark energy EoS $w$, these data combinations give consistent results on $H_0$ and $\Omega_m$, which is listed in table \ref{wcdmtable}. And the minimal $\chi^2$ in the $w$CDM model is much smaller than that in the standard $\Lambda$CDM model, $\Delta\chi^2_{\rm min}=-8.82$. The difference on the AIC is $\Delta{\rm AIC}=-6.82$ as a consequence. The tension between constraints on $H_0$ almost disappears in the $w$CDM model and the constraints of $w$ imply that the current data still favor the dark energy model with $w<-1$ strongly.

Finally, we add the BAO information into our analysis. From Table \ref{wcdmtable} we can see that in the $\Lambda$CDM model, the BAO data could also significantly improve the constraints on $H_0$ by a factor of $1.5$ and $2.3$ when comparing with those from CMB data alone, namely $H_0=68.30\pm0.96$ ${\rm km\,s^{-1}\,Mpc^{-1}}$ and $H_0=67.69\pm0.79$ ${\rm km\,s^{-1}\,Mpc^{-1}}$ ($68\%$ C.L.) from WMAP9+BAO and Planck+WP+BAO. In the right upper panel of Figure \ref{wcdm} we show the 2-dimensional contour in the $(H_0,\Omega_m)$ panel in the $\Lambda$CDM. Differing from the above cases, WMAP9+BAO and Planck+WP+BAO give the similar constraint on $H_0$ and $\Omega_m$. However, their constraints are still apparently different from the HST prior on $H_0=73.8\pm2.4$ ${\rm km\,s^{-1}\,Mpc^{-1}}$ at about $7.7\,\sigma$ confidence level. This would imply that the BAO information also favor a low value of $H_0$ in the $\Lambda$CDM model \cite{planck_fit}. When the dark energy EoS $w$ is a free parameter, the obtained median value of $H_0$ from Planck+WP+BAO becomes larger and consistent with that of the HST prior.

Based on these analyses, the tensions between constraints on $H_0$ between Planck+WP and WMAP9 or the HST prior could be eased by relaxing the assumptions about the dark energy model. As we know, $w$ is anti-correlated with $H_0$. The $\Lambda$CDM model forces the EoS of dark energy to be $w=-1$, which strongly limits the allowed parameter space of $H_0$ and obtains a low value of $H_0$. When we perform this analysis in the $w$CDM framework, the value $w$ can be less than $-1$, such that a higher $H_0$ and a lower $\Omega_m$ can provide the same fit to the data. This $H_0$ tension disappears. Overall, the tension between constraints on $H_0$ found by Planck+WP actually gives a hint that the Planck data favor the dark energy model with $w<-1$. If we allow the EoS of dark energy to be smaller than $-1$, the Planck+WP data will give a consistent result on $H_0$ with that from WMAP9 or the HST gaussian prior.

\section{Dark Energy with Time-evolving $w$}\label{quintom}

In this section, we extend our analysis to the dark energy model with the parameterized time-evolving EoS [Eq. (\ref{eq_cpl})]. In Table \ref{cpltable} we present the constraints on the dark energy parameters $w_0$, $w_a$ and some other parameters from different data combinations.

\begin{table}
\caption{The median values and $1\,\sigma$ error bars on some cosmological parameters obtained from different data combinations.}\label{cpltable}
\begin{center}

\begin{tabular}{c|c|c|c|c}

\hline\hline
Parameters&$w_0$&$w_a$&$H_0$&$\Omega_m$\\
\hline
Planck+WP+BAO&$-0.79\pm0.57$&$-1.1\pm1.7$&$68.2\pm5.5$&$0.315\pm0.052$\\
Planck+WP+SNLS&$-0.81\pm0.19$&$-1.9\pm1.1$&$73.6\pm2.4$&$0.264\pm0.019$\\
Planck+WP+HST&$-0.10\pm0.66$&$-5.3\pm3.1$&$72.7\pm2.7$&$0.270\pm0.021$\\
Planck+WP+SNLS+BAO&$-1.08\pm0.13$&$-0.13\pm0.53$&$70.7\pm1.6$&$0.285\pm0.012$\\
WMAP9+SNLS+BAO&$-1.07\pm0.16$&$-0.19\pm0.86$&$70.3\pm1.7$&$0.285\pm0.014$\\
\hline\hline
\end{tabular}
\end{center}
\end{table}

\subsection{Dark Energy Perturbations}\label{perturbation}

The consistence for the global analysis requires the inclusion of the dark energy perturbations. However, the perturbation diverges when the w crosses the cosmological constant boundary. This can be seen explicitly from the following equations \cite{pert_ma}:
\begin{eqnarray}
\dot{\delta}&=&-(1+w)(\theta-3\dot{\Phi})-3\mathcal{H}(c^2_s-w)\delta~,\\
\dot{\theta}&=&-\mathcal{H}(1-3w)\theta-\frac{\dot{w}}{1+w}\theta+k^2\left(\frac{c^2_s\delta}{1+w}+\Psi\right)~.
\end{eqnarray}
Mathematically it has been proved that the single-field dark energy models like quintessence \cite{quintessence}, phantom \cite{phantom} and k-essence \cite{kessence} cannot give rise to w crossing -1. To have the w across the cosmological constant boundary, it needs the Quintom model \cite{quintom} where the extra degree of freedom have been introduced. In the literature, there have been many attempts to build models of this class and we refer to Ref.\cite{cai} for a recent review.

With the parametrization of the equation of state in Eq.(\ref{eq_cpl}), the dark energy perturbation needs to be carefully treated. In 2005, we made a proposal \cite{pert_zhao, pert_xia} based on the Quintom theory. Technically, we introduce a small positive constant $\epsilon$ to divide the full range of the allowed value of w into three parts: (1) $w > -1 + \epsilon$; (2) $-1 - \epsilon \leq w \leq -1 + \epsilon$; and (3) $w < -1 - \epsilon$. Neglecting the entropy perturbation contributions, for the regions (1) and (3) the equation of state does not cross the $w\equiv-1$ boundary and perturbations are well defined by solving the above equations. For the region (2), we set $\dot{\delta}=0$ and $\dot{\theta}=0$ and match the perturbation in region (2) to those of the regions (1) and (3) at the boundary. In the numerical calculations we limit the range to $|\Delta w = \epsilon| < 10^{-5}$ and find that this method is a very good approximation to the realistic model of dark energy which gives rise to w across -1.

\begin{figure}[t]
\begin{center}
\includegraphics[scale=0.6]{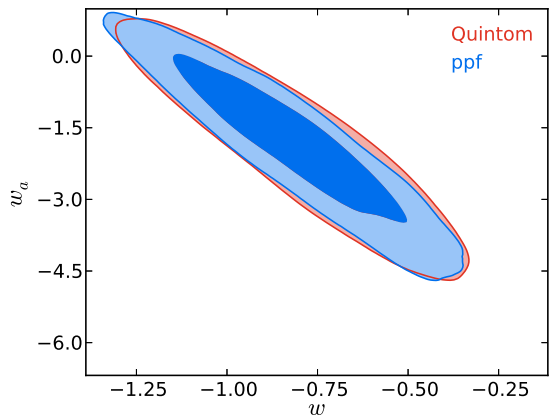}
\caption{Marginalized one-dimensional and two-dimensional likelihood (1, $2\,\sigma$ contours) constraints on the parameters $w_0$ and $w_a$ from Planck+WP+SNLS with different method for treating perturbations of time evolving EoS of dark energy: the Quintom method (red) and the PPF method (blue). \label{cpl_check}}
\end{center}
\end{figure}

In the Planck analysis \cite{planck_fit}, they use the parameterized post-Friedmann (PPF) method to handle the dark energy perturbations when w crosses $-1$, which was proposed by the Ref.\cite{pert_fang} in the year of 2008. We have checked the consistency between these two methods of treating the dark energy perturbation, which is shown in Figure \ref{cpl_check}.

Since the CMB data alone can not constrain the time-evolving dark energy model very well, here we use the Planck+WP data and the SNLS data together to do the global analysis, in order to obtain the meaningful constraints. When using our own method, we obtain the constraints: $w_0=-0.81\pm0.19$ and $w_a=-1.85\pm1.1$ at $68\%$ confidence level. If we use the PPF method to handle the dark energy perturbations, the data sets yield the almost identical limits on dark energy parameters of $w_0=-0.81\pm0.19$ and $w_a=-1.84\pm1.1$ ($68\%$ C.L.). We plot the 2-dimensional contours in the ($w_0,w_a$) panels obtained from these two methods in Figure \ref{cpl_check}. As can be seen from this plot, the two methods are equivalent. In this paper, we will use our own method for the dark energy perturbations in the numerical calculations.

\subsection{Constraints on $w_0, w_a$ }

\begin{figure}[t]
\begin{center}
\includegraphics[scale=0.6]{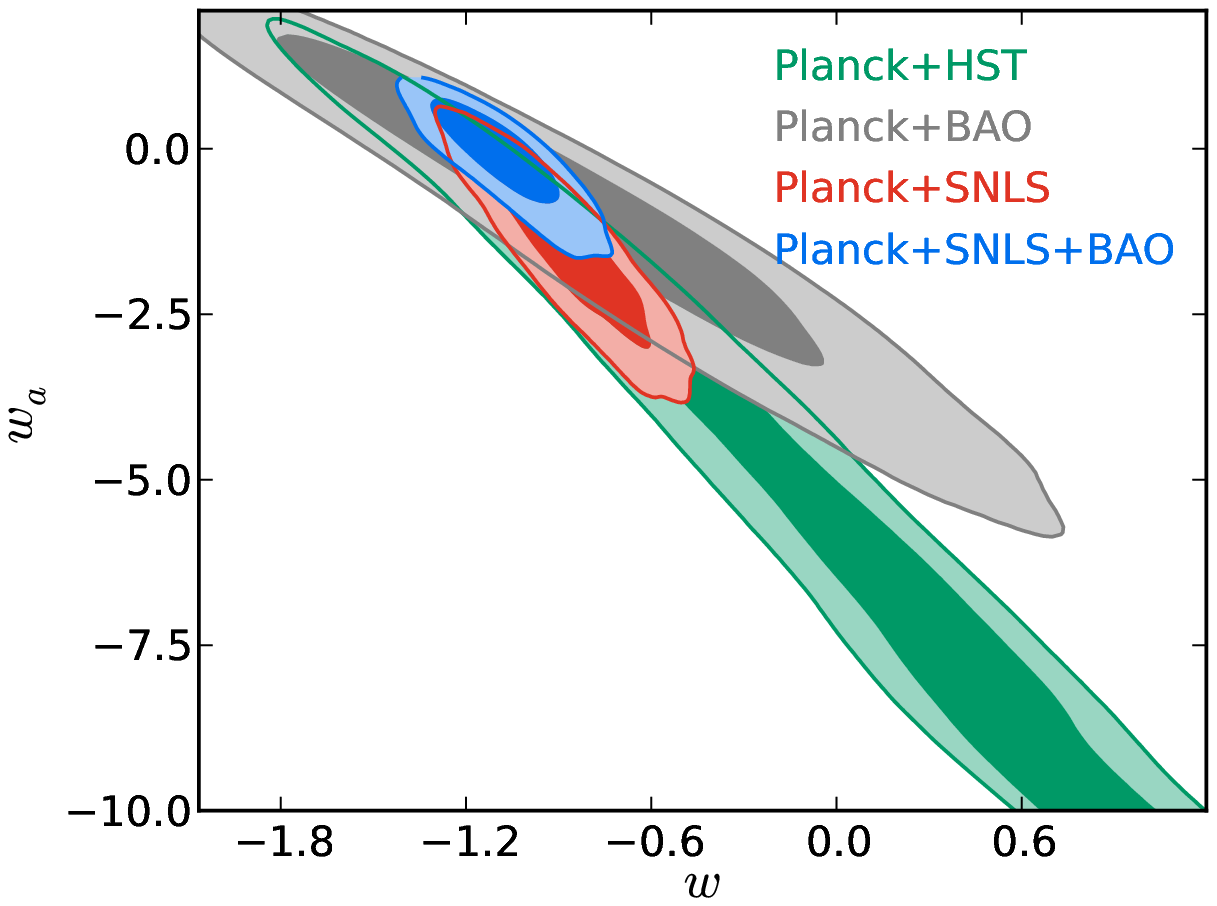}
\includegraphics[scale=0.41]{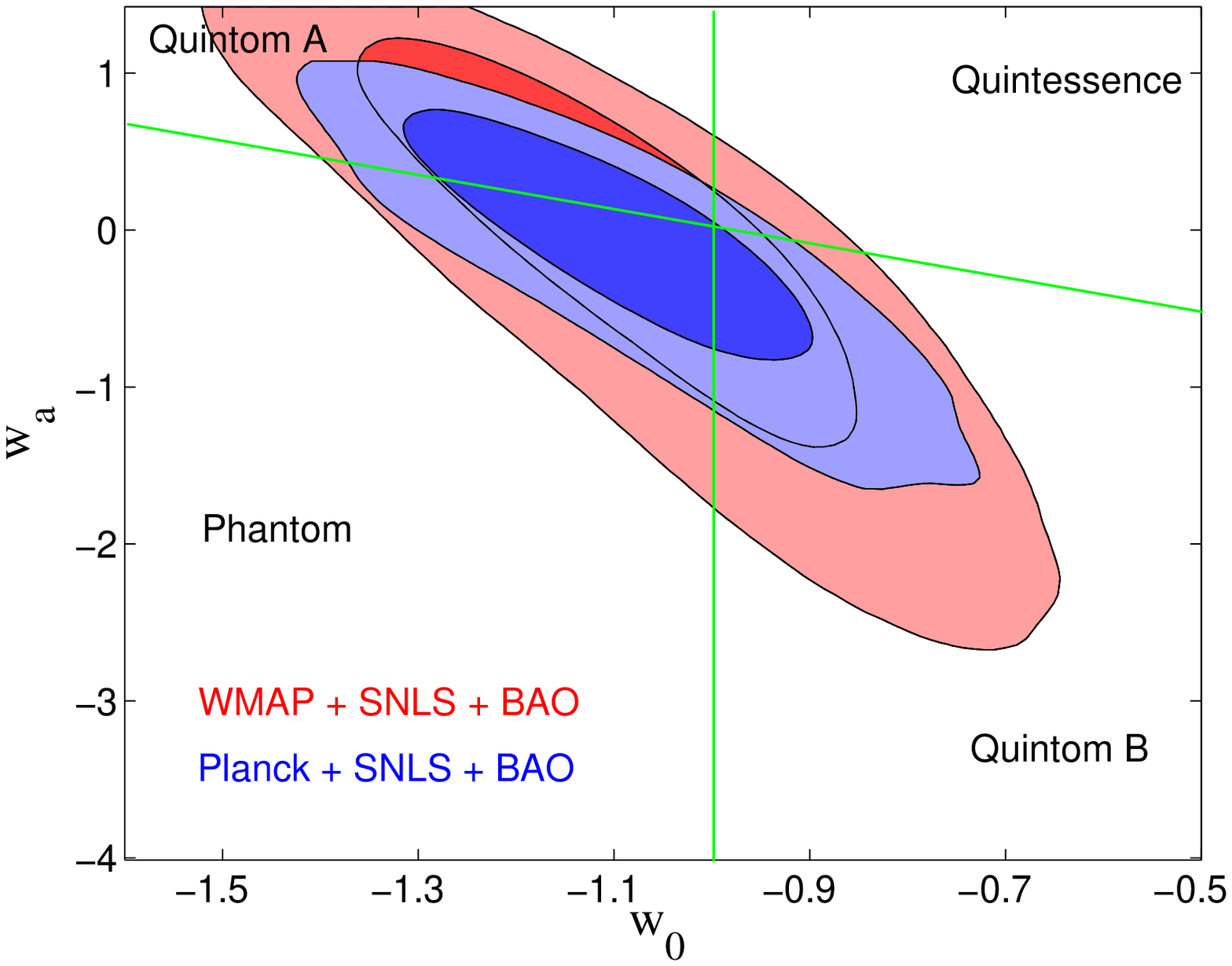}
\caption{Marginalized two-dimensional likelihood (1, $2\,\sigma$ contours) constraints on the time-evolving EoS of dark energy from different data combination. The green solid lines stand for $w_0=-1$ and $w_0 + w_a = -1$. \label{cpl_plk_difd}}
\end{center}
\end{figure}

In this subsection we present the constraints on the dark energy parameters from different data combinations and show the 2-dimensional contours in the ($w_0,w_a$) panel in Figure \ref{cpl_plk_difd}. We first use the Planck+WP+SNLS data combination. The $68\%$ constraints on $w_0$ and $w_a$ are $w_0=-0.864\pm0.17$ and $w_a=-1.54\pm0.92$, which implies that the dynamical dark energy models are not excluded \cite{Xia:2008ex,Li_2013}. When comparing with the standard $\Lambda$CDM model, the minimal $\chi^2$ in the $w(z)$CDM is significantly smaller, $\Delta\chi^2_{\rm min}=-10.62$, and the value of AIC becomes also smaller, $\Delta{\rm AIC}=-6.62$. The data favor the dynamical dark energy model with an equation of state getting across $w=-1$ during the evolution of the Universe \cite{quintom}, and the standard $\Lambda$CDM model is not favored by the data at about $2\,\sigma$ confidence level. Combining the Planck+WP and the HST prior gives the similar results. Since the constraining power of HST prior is much less than that of the SNLS data, the constraints on $w_0$ and $w_a$ are much weaker, namely the $95\%$ C.L. constraints are $w_0=-0.10\pm0.66$ and $w_a=-5.2\pm3.1$. However, in this data combination, we obtain $\Delta\chi^2_{\rm min}=-7.56$ and $\Delta{\rm AIC}=-3.56$, when comparing the constraints in the standard $\Lambda$CDM and the $w(z)$CDM models. In Figure \ref{cpl_plk_difd}, we can also see that the standard $\Lambda$CDM model is not favored by the data at about $2\,\sigma$ confidence level.  In these two cases, we obtain the similar constraints on $H_0$: $H_0=73.6\pm2.4$ ${\rm km\,s^{-1}\,Mpc^{-1}}$ and $H_0=72.7\pm2.7$ ${\rm km\,s^{-1}\,Mpc^{-1}}$ ($68\%$ C.L.) from Planck+WP+SNLS and Planck+WP+HST, respectively. Clearly there is no tension on the Hubble constant in the dynamical dark energy model.

When we use the Planck+WP+BAO data sets, the direction of 2-dimensional contour in the ($w_0,w_a$) panel is different from that in the above two cases, see Figure \ref{cpl_plk_difd}. And the BAO information favors a smaller value of $w_0$ and a higher value of $w_a$: $w_0=-0.79\pm0.57$ and $w_a=-1.1\pm1.7$ ($68\%$ C.L.). Therefore, when combining Planck+WP, SNLS and BAO together, the contour for the combined dataset are tighter than that obtained from Planck+WP+SNLS. The standard $\Lambda$CDM model falls within $2\,\sigma$ confidence region, which can be seen from the left panel of Figure \ref{cpl_plk_difd}. The $95\%$ C.L. constraints on $w_0$ and $w_1$ are $-1.3 < w_0 < -0.81$ and $-1.3 < w_a < 0.81$. We also do the calculation by replacing the Planck+WP by the WMAP9 data and find that the constraints from WMAP9+SNLS+BAO are similar to that obtained from Planck+WP+SNLS+BAO, but with larger error bars, namely $-1.4 < w_0 < -0.72$ and $-2.0 < w_a < 1.3$ ($95\%$ C.L.), as shown in the right panel of Figure \ref{cpl_plk_difd}. In conclusion, these results imply that the constraints on dark energy parameters are dependent on the cosmological data sets we use \cite{Zhao:2012aw,Li:2011dr,Appleby:2013upa}, since the current data sets are not in good agreement.

\section{Summary}\label{sum}

Recently, the Planck collaboration published the cosmological results from the highest resolution, full sky, maps of the CMB temperature anisotropies. In the standard $\Lambda$CDM model, the Planck+WP data give a lower value of $H_0$ when comparing with some low redshift probes, such as the direct $H_0$ probes. This tension between constraints on $H_0$ has been discussed in the literature \cite{verde,waynehu,huang} and probably can be resolved in the dark energy model with a non-trivial EoS. When using a constant EoS dark energy model, we find that the $H_0$ measurements from the CMB are highly model-dependent and the value of $H_0$ can be changed significantly. The constraint on $H_0$ from Planck+WP data is now consistent with the HST gaussian prior from the local probe when we extend the standard $\Lambda$CDM to the $w$CDM model. This $H_0$ tension found by Planck+WP actually implies that the Planck data favor the dynamical dark energy model.

Then, we constrain the dark energy model with a time-evolving EoS. Since the dark energy perturbations are crucial in the analysis, we compare our own method with the PPF method proposed by Ref.\cite{pert_zhao} in handling the dark energy perturbations when the EoS gets across $w\equiv-1$. From the data combination Planck+WP+SNLS, we use both methods to do the analyses and obtain almost identical constraints on the dark energy parameters. Based on our method, we find that the $\Lambda$CDM model is disfavored at about $2\,\sigma$ C.L. from the data Planck+WP+SNLS and Planck+WP+HST. The minimal $\chi^2$ and the value of AIC in the $w(z)$CDM model have been significantly reduced. The current cosmological data slightly favors the dynamical dark energy models and the best fit model is the Quintom scenario whose EoS crosses the cosmological constant boundary $w\equiv-1$, which are also consistent with those obtained from the information of CMB distance priors \cite{Li:2008cj,Li:2010yb,Wang:2013mha}.


\section*{Acknowledgements}

We acknowledge the use of the Legacy Archive for Microwave Background Data  Analysis (LAMBDA). Support for LAMBDA is provided by the NASA Office of Space Science. JX is supported by the National Youth Thousand Talents Program and the grants No. Y25155E0U1 and No. Y3291740S3. HL is supported in part by the National Science Foundation of China under Grant Nos. 11033005, by the 973 program under Grant No. 2010CB83300, by the Chinese Academy of Science under Grant No. KJCX2-EW-W01. XZ is supported in part by the National Science Foundation of China under Grants No. 10975142 and 11033005, and by the Chinese Academy of Sciences under Grant No. KJCX3-SYW-N2.


\end{document}